%% file: pdketa.tex
\documentstyle[preprint,eqsecnum,tighten,floats,epsf,12pt,prl,aps]{revtex}

\include{ltkmcrs}

\title{ 
{
\begin{flushright}
{\normalsize 
TUHEP-2000-01\\
PDK-744\\}
\end{flushright}
\bf Search for Nucleon Decay with Final States \\
 $\ell^+\eta^0$, $\overline{\nu}\eta^0$, and $\overline{\nu}\pi^{+,0}$ Using
 Soudan 2}
      }
      
\author{
D.~Wall$^{5,a}$, 
W.W.M.~Allison$^3$, G.J.~Alner$^4$, D.S.~Ayres$^1$, W.L.~Barrett$^6$,\\ 
C.~Bode$^2$, P.M.~Border$^2$, J.H.~Cobb$^3$,
R.~Cotton$^4$, H.~Courant$^2$, J.~Chung$^5$,\\
D.M.~Demuth$^2$, T.H.~Fields$^1$, H.R.~Gallagher$^3$, M.C.~Goodman$^1$,
 R.~Gran$^2$, \\ T.~Joffe-Minor$^1$, T.~Kafka$^5$, S.M.S.~Kasahara$^2$, 
W.~Leeson$^1$,  P.J.~Litchfield$^4$, \\ W.A.~Mann$^5$, M.L.~Marshak$^2$,
R.H.~Milburn$^5$, W.H.~Miller$^2$,  L.~Mualem$^2$, \\ A.~Napier$^5$,
 W.P.~Oliver$^5$, G.F.~Pearce$^4$, E.A.~Peterson$^2$, D.A.~Petyt$^4$,\\
L.E.~Price$^1$, K.~Ruddick$^2$, M.~Sanchez$^5$, J.~Schneps$^5$,
 A.~Stassinakis$^3$,\\ J.L.~Thron$^1$, S.P.~Wakely$^2$, N.~West$^3$\\
\vspace{.15in}
$^1${\it Argonne National Laboratory, Argonne, IL 60439}\\
$^2${\it University of Minnesota, Minneapolis, MN 55455}\\
$^3${\it Department of Physics, University of Oxford, Oxford OX1 3RH, UK}\\
$^4${\it Rutherford Appleton Laboratory, Chilton, Didcot, Oxfordshire
 OX11 0QX, UK}\\
$^5${\it Tufts University, Medford, MA 02155}\\
$^6${\it Western Washington University, Bellingham, WA 98225}\\
\vspace{.15in}
$^a${Now at Sapient Corporation, One Memorial Drive, Cambridge, MA 02142}\\}

\begin{document}


\maketitle 
\thispagestyle{empty}

\begin{abstract}

We have searched for nucleon decay into five two--body final states
using a 4.4 kiloton-year fiducial
exposure of the Soudan 2 iron tracking calorimeter. For proton decay into the
fully visible final states $\mu^+\eta^0$ and e$^+\eta^0$, we observe zero and
one event, respectively, that satisfy our search criteria for nucleon decay.
The lifetime lower limits
($\tau /B$) thus implied are $89 \times 10^{30}$ years and
$81 \times 10^{30}$ years at 90\% confidence level. 
For neutron decay into $\overline{\nu}\eta^0$, we obtain the lifetime lower
limit $71\times10^{30}$ years. Limits are also
reported for neutron decay into $\overline{\nu}\pi^0$, and for proton decay into
$\overline{\nu}\pi^+$. 


\end{abstract}

\eject

\section{Introduction}

\subsection{SUSY predictions for nucleon decay}

The decay of the nucleon is a possible experimental
window into fundamental processes at high mass scales.
Calculations carried out in supersymmetric (SUSY) grand unified theories (GUTs),
such as SUSY SU(5) and SUSY SO(10), have become increasingly more 
detailed, and predicted lifetimes are within a range which may be accessible
to experiment. In contrast to their non-SUSY precursors, SUSY GUT models
generally favor decay modes with final state K$^0$ or K$^+$ mesons.
Some of these models also predict that other decay modes may exhibit a
significant branching fraction. The latter modes include the lepton plus
eta meson and the $\overline{\nu} \pi$ modes.
In SUSY SU(5) models, lifetimes predicted for $\ell^+\eta$ and
$\overline{\nu}\eta$ are longer than those for $\overline{\nu}$K or
$\overline{\nu}\pi$ modes by factors which vary from 2-3 to one or two orders
of magnitude\cite
{theory:nath_and_chamseddine_and_arnowitt,theory:Hisano_and_Murayama,Nath2}.
Eta mode lifetimes which are hundreds of times longer than
$\overline\nu$K or $\ell^+$K
lifetimes have also been predicted in SUSY SO(10)
\cite{Macpherson,theory:lucas_and_rabi}.
However, it has been proposed \cite{babu_pati,Pati}
 that in SUSY GUTs such as SO(10), there exists
a new set of color triplets and thereby a new source of $d$=5
operators which may allow p $\rightarrow \ell^+\eta$  to become prominent
(along with $\overline\nu$K$^+$ and $\overline\nu\pi^+$).
In SUSY GUTs the $\overline{\nu}\pi$
modes are usually predicted to have decay lifetimes which
are longer than $\overline{\nu} K$ modes.
In specific models, however, it is possible for interference between third
and second generation amplitudes to alter this situation so that
$\overline{\nu} \pi$ modes become dominant
\cite{theory:nath_and_chamseddine_and_arnowitt,Nath2}.

We have previously searched for nucleon decay final states with
strangeness using the Soudan 2 iron tracking calorimeter and reported
the results as lifetime lower limits \cite{Soudan_nuK+,DWall}.
In this work, we extend our investigations to include the $\ell^+\eta^0$,
$\overline{\nu}\eta^0$, $\overline{\nu}\pi^0$, and $\overline{\nu}\pi^+$
final states.

\subsection{Previous searches for $\ell^+\eta,\ \overline{\nu}\eta^0,$ and
$\overline{\nu}\pi$ modes}

Two-body $\ell^{+,0}\eta$ and $\overline{\nu}\pi$ decays of nucleons were
regarded as interesting in the context of non-SUSY GUT models, and
experimental searches were reported at various times during the past two
decades. Multiple ring topologies were investigated for $\ell^+ \eta$,
$\overline{\nu} \eta$, and $\overline{\nu} \pi^0$ final states by the IMB and
Kamiokande water Cherenkov experiments \cite{IMB,IMB2,Kamiokande}.
In the HPW water Cherenkov experiment, events
exhibiting a two muon decay signature were examined for compatibility 
with p $\rightarrow \mu^+ \eta$, with $\eta$ decaying to $\pi^+ \pi^- \pi^0$
or $\pi^+ \pi^- \gamma$ \cite{HPW}.
The $\ell^+ \eta$, $\overline{\nu} \eta$, and
$\overline{\nu} \pi$ modes were also investigated using 
the planar iron tracking calorimeter of Fr\' ejus \cite{Frejus,Frejus2}.   
More recently, substantial improvement of lifetime limits for the two-body
eta modes has been reported, based upon the 7.6 kton-year
exposure of the IMB-3 water Cherenkov detector \cite{IMB-3}. Generally,
these experiments observed zero or modest numbers of candidate events.
Understandably, the more loosely constrained
n $\rightarrow \overline{\nu} \eta$ and  n $\rightarrow \overline{\nu} \pi^0$
modes were found to exhibit more background and consequently to
yield less stringent lifetime lower limits.

\section{Detector and event samples}

\subsection{Central Detector and Active Shield}

     Soudan 2 is a massive 963 (809 fiducial) metric-ton iron 
tracking calorimeter.  It operates as a slow-drift, fine-grained,
time projection chamber of honeycomb-lattice geometry
and with $dE/dx$ imaging of tracks and showers.  As such, it
differs considerably from water Cherenkov detectors and from
planar iron tracking calorimeters.

      The construction and performance of the Soudan calorimeter
are described in previous publications \cite{Soudan_NIM_1}. In brief,
charged particles are imaged via one meter long, slightly
conductive, plastic drift tubes.  Electrons liberated by
throughgoing charged particles drift to the tube ends under the action of
a voltage gradient applied along the
tubes.  The tubes are sandwiched between mylar sheets so as to 
comprise a ``bandolier" assembly. 
Corrugated steel sheets with interleaved bandolier are then
stacked to form a massive lattice.  A stack is packaged with wireplanes
and cathode pickup strips at the drift tube ends and surrounded
by thin steel skins which provide the gas enclosure.  The resulting
assemblies are $1 \times 1 \times 2.5$ m, 4.3 ton calorimeter
modules.  The tracking calorimeter is constructed building-block
fashion, with contiguous ``walls" which are two modules high and eight modules
wide.
The calorimeter is surrounded on all sides by a double-layer, cavern-liner
proportional tube active shield of 1700 m$^2$ total tracking area
\cite{Soudan_NIM_2}. Shield augmentations in the form of additional single
layers and a double-layer top cover of proportional tubes have been
operational since 1995.

For the task of searching for nucleon decay in a variety of final states,
the Soudan 2 calorimeter provides imaging of non-relativistic as well as
relativistic tracks, and its systematics are different from those of
other experiments.  The detector is located at a depth of
713 meters (2070 mwe) in the Tower-Soudan Underground Mine State Park in
northern Minnesota. Data-taking is still underway, having commenced in
April 1989 when the total mass was 275 tons, and continued
as more modules were installed.  Operation with the 
calorimeter at full mass was first achieved in November 1993.   The analysis
reported in this work is based on a total (fiducial) exposure of
5.52 (4.41) kiloton-years, obtained with data taken through
October 1998.

\subsection{Data and Monte Carlo event samples}
\label{sec:twob}

Four distinct event samples have been assembled for this analysis;  two of
the samples are generated as Monte Carlo (MC) simulations with full detector
response, and two are comprised of data events recorded in the experiment.
These samples, details of which have been presented elsewhere
\cite{Soudan_nuK+,DWall}, are:

\newcounter{bean}
\begin{list}
{\it\roman{bean})}{\usecounter{bean}
\setlength{\rightmargin}{\leftmargin}}
\item {\it Nucleon decay MC}:
For each of the nucleon decay processes investigated, we use
MC simulations which track all final state particles
through the detector geometry. Electronic hits are generated,
and detector noise is included, in the same format as with real data.
\item {\it Atmospheric neutrino MC}:
Events are generated representing charged current and
neutral current reactions which are initiated by the flux of
atmospheric neutrinos.  The neutrino MC program is based upon the 
flux calculation of the Bartol group for the Soudan site \cite{Gaisser}.
The neutrino MC sample analyzed here corresponds to an exposure of
24.0 fiducial kton-years.
\item {\it Rock data}:
We analyze events for which the veto shield recorded
       coincident, double-layer hits.  Such events originate with inelastic
       cosmic-ray muon interactions in the cavern rock surrounding the
       detector.  These shield-tagged ``rock" events provide a reference
       sample by which to gauge cosmic-ray induced background events which
       are included in the ``gold data".  The latter events arise either
       from shield inefficiency or from instances where an energetic
       neutral particle emerged from the cavern walls with no accompanying
       charged particles.
\item {\it Gold data}:
Data events for which the cavern-liner active shield array was
       quiet during the allowed time window comprise our
       ``gold event" sample.   These events are mostly reactions initiated
       by atmospheric neutrinos but may contain nucleon decays, as well as
the muon induced rock events with no shield hits (described above).
The events of interest for our study are those with a ``multiprong" topology,
having two or more particles (other than recoil nucleons) emerging from the
primary vertex. They are distinct from the more populous single-track and
single-shower events which originate predominantly from neutrino
quasi-elastic reactions.
\end{list}
   
Events of all four samples are required to be fully contained in a fiducial
volume which is everywhere 20 cm or more from the outer surfaces of the
calorimeter. In order for an event to be included in any of the above four
samples it must survive the selections of a standard processing chain.  
At the head of the chain are the requirements imposed by the hardware
trigger; events satisfying these requirements are subjected to
a containment filter code.  Events which survive are subjected to
two successive scanning passes carried out by physicists; each pass
involves three complete, independent scans. In the first scan pass, events
with multiprong topologies used in this study are found with an overall
efficiency of $0.98^{+0.02}_{-0.04}$. Descriptions of our
hardware and software selections, and of our scanning procedures are
given elsewhere \cite{Soudan_nuK+,DWall,Soudan_ratio_97}. Our procedures 
ensure that MC simulation events pass through identical
or otherwise equivalent (e.g. the hardware trigger is implemented via
software for MC events) steps in the chain \cite{DWall}.

Events which emerge from the second scan pass with a multiprong
topology assignment are then reconstructed using an interactive
graphics package.   The set of reconstructed tracks and showers which
comprises each event is subsequently entered into an event summary file
from which kinematic quantities are calculated.

\subsection{Background from rock events and from neutrino interactions}
 
      Among the fully contained, multiprong events of our gold data
sample, a small contribution may arise from cosmic ray induced rock events.
These are events initiated by neutrons emerging from the cavern rock which
impinge upon the central detector and which are unaccompanied by coincident
hits in the surrounding shield array.   In contrast to neutrino 
interactions or to nucleon decay, these shield-quiet rock events
tend to occur at relatively shallow penetration depths into the calorimeter.
Their depth distribution, and their distribution in visible energy and in
other variables, can be inferred from rock events tagged by coincident
shield hits.   To estimate the amount of zero-shield-hit rock background
in our gold multiprong sample, event distributions of gold data have been
fitted to neutrino MC and shield-tagged rock samples using a multivariate 
discriminant analysis \cite{DWall,Leeson:PDK684}. 
We find that, the fraction $f$ of all multiprong rock events which have
at least one shield hit is $f = 0.94 \pm 0.04$.

     For each individual nucleon decay channel, the same event selections  
applied to data multiprongs are also applied to the shield-tagged rock
multiprongs.  The zero-shield-hit rock background is then estimated as
the product $(1-f)/f = 0.064$ times the number of shield-tagged rock
events which satisfy the channel selections.

      To calculate rates for background events in our nucleon decay search
which arise from interactions of atmospheric neutrinos in the detector,
we use our realistic neutrino MC simulation which is based
upon atmospheric fluxes with null oscillations.  The neutrino Monte Carlo
program has been described in previous publications 
\cite{DWall,Soudan_ratio_97}.

      During the past decade, evidence for depletion of the atmospheric 
muon-neutrino flux as described by $\nu_\mu \rightarrow \nu_x$ oscillations
has become increasingly extensive; especially compelling are observations by 
Super-Kamiokande of zenith angle distortions in fluxes of both 
sub-GeV and multi-GeV muon neutrinos \cite{SK}. The disappearance 
of $\nu_\mu$ flavor neutrinos by oscillations effectively reduces background
in our search arising from ($\nu_\mu + \overline\nu_\mu$) 
charged-current reactions, and so a correction for this effect to
our null oscillation estimates is warranted.  To implement a 
correction we assume, as indicated by recent data, that $\nu_x$ is an
active neutrino which is not $\nu_e$ (i.e. $\nu_x = \nu_\tau$) \cite{LP99}.
Then, atmospheric neutrino oscillations do not affect background arising
from ($\nu_e + \overline\nu_e$) charged-current reactions, nor do they affect
background from neutral current reactions (initiated by any flavor).
Consequently, to correct for $\nu_\mu$ flavor disappearance, we simply scale
the number of ($\nu_\mu + \overline\nu_\mu$) charged-current background events
estimated from the null oscillation MC for each nucleon decay channel by
the $\nu_\mu/\nu_e$ flavor ratio $R$ measured in the Soudan 2 experiment:
$R = 0.64 \pm 0.13$ \cite{Soudan_ratio_99}.
As noted below in Sections IV and V, and in Table II,
this correction yields small reductions in null oscillation background 
rates for p $\rightarrow \mu^+ \eta$ and p $\rightarrow \overline\nu \pi^+$
final states. For p $\rightarrow {\rm e}^+ \eta$ and
n $\rightarrow \overline\nu \pi^0$ channels however, the neutrino oscillation
correction has negligible effect.

\section{Nucleon decay simulation and search method}
\label{sec:three}

\subsection{Event generation}

For each nucleon decay final state, a Monte Carlo sample is created and
processed as described in Section~\ref{sec:twob}. This sample is then used
to determine the topological and kinematic properties that differentiate it
from the atmospheric neutrino and the rock event backgrounds.

For each final state, about 500 events are generated and embedded into pulser
trigger events from the detector which are taken at regular intervals
throughout the exposure.   In this way, the detector's evolving size
and the background from natural radioactivity and 
cosmic rays are accurately incorporated into the simulation.  

For two-body decay of a nucleon at rest, the final state momenta
are uniquely determined.  However, Fermi motion within parent iron and
other nuclei of the calorimeter medium smears these momenta and thereby
complicates final state identification.  In our simulations, Fermi motion
effects are modelled according to the parameterization of
Ref.~\cite{theory:bodandrit}. 

A Monte Carlo event for p $\rightarrow \mu \eta$,
$\eta \rightarrow \gamma \gamma$, which illustrates the search topology for
this mode, is shown in Fig.~\ref{fig:fig1}.
Here, the $\mu^+$ appears as a single non-scattering
track.  The two gammas from $\eta$ decay give rise to two showers which
are spatially well separated and which point to the event vertex which
is also the origin of the muon.  The $\mu^+$ endpoint decay -  discernible in
approximately 60\% of events - appears in Fig.~\ref{fig:fig1} as extra
ionization ``hits" (tube crossings) in the vicinity of the track's range-out
point. 

      In each nucleon decay mode, the meson can undergo
intranuclear rescattering within the parent nucleus.  For nuclei which
have interior as well as surface nucleons ($A \ge 12$), there is significant
probability for event final states to be altered.   For final state
pions (in neutrino MC interactions as well as
in nucleon decay), intranuclear rescattering is treated using a
phenomenological cascade model \cite{mann:intranuke:leeson:pdk678}. Parameters
of the model were set by requiring that the threshold pion production observed
in $\nu_{\mu}$--deuteron ($A=2$) and $\nu_{\mu}$--neon ($A=20$) reactions
be reproduced \cite{merenyi}.
To account for the intranuclear rescattering of $\eta$
mesons within iron nuclei we have adopted the survival fraction of
0.57 estimated by the Fr\'{e}jus  collaboration \cite{Frejus} using a
detailed balance calculation. This value for the $\eta^0$ survival is
similar to the survival fraction of 0.52 which we calculate for the
$\pi^0$ of n $\rightarrow \overline{\nu}\pi^0$ in iron nuclei.
We note that a recent analysis of $\eta$ meson photoproduction in nuclei finds
the in-medium $\eta$N cross-section to be nearly independent of $\eta$
momentum between 0 and 500 MeV/$c$\cite{Effenberger}. 

\subsection{{\bf Search contour in the} $\bbox{M_{inv}}$ {\bf versus}
$\bbox{\left| \vec{p}_{net} \right|}$ {\bf plane}}
\label{sec:threec}

Two quantities that are useful for selection of nucleon decay candidates
and the rejection of backgrounds are the invariant mass,
$M_{inv}$, and the magnitude of the net three--momentum,
$\left| \vec{p}_{net} \right|$, of the reconstructed final state.

We create a scatter plot of invariant mass versus net event momentum for
the reconstructed final states for each simulation. Then we choose a region
in this plane whose boundary defines a kinematical selection which can be
applied to the data and to the background samples. We observe that, for
most nucleon decay final states, event distributions in each of
these variables are approximately Gaussian. Consequently, the density
distribution of points on the invariant mass versus momentum plane can be
well represented by a bi--variate Gaussian probability distribution function.
A detailed description of this construction is given elsewhere
\cite{DWall,DWall_thesis}.

      Projections of bi--variate Gaussian surfaces onto 
the $M_{inv}$ versus $\left| \vec{p}_{net} \right|$ plane for the eta modes
of this search are shown in Figs. 2 -- 7. In these figures, 
the distribution of a nucleon decay event population
is depicted using five nested, elliptical boundaries. Proceeding
outward from the innermost contour, the bounded regions contain respectively
10\%, 30\%, 50\%, 70\%, and 90\% of the MC nucleon decay sample.
From the five regions
delineated we choose the 90\%-of-sample contour - the outermost, solid-curve
ellipse in Figs. 2--7 - to define our ``primary" kinematic
selection. That is, a candidate nucleon decay event has reconstructed
($M_{inv}$,  $\left| \vec{p}_{net} \right|$) values which
lie within the outer contour.  The interior
contours (dashed ellipses in Figs. 2--7) are helpful to gauge
whether an event sample as a whole exhibits the kinematics of
nucleon decay, and for this reason we display them. For our search
involving the $\eta \rightarrow \gamma\gamma$ decay, containment within
the 90\% contour is our sole kinematic constraint; no subsidiary condition,
such as a cut on the $\gamma\gamma$ invariant mass, has been used.

All nucleon decay simulation events are subjected to the triggering
requirements, the detector containment requirements, and to the scanning and
topology cuts. Subsequently, kinematic cuts and additional topology
cuts, designed to reduce the background on a mode--by--mode basis, are applied.
The cumulative effect of these selections is to reduce
the overall detection efficiency significantly. Typically, the product of the
triggering, containment, and scanning selections reduces the
survival fraction to below 30\% for any mode.
Table~\ref{tbl:efficiencies_noK} shows the successive survival fractions
(including $\eta$ decay branching ratios) for all modes studied.
Note that the effects of intranuclear rescattering processes (INR) are included
in the survival fractions listed in Table~\ref{tbl:efficiencies_noK}.
For comparison, the rightmost column of  Table~\ref{tbl:efficiencies_noK}
shows our estimates of $\epsilon\times$BR in the absence of INR within the
iron and lighter nuclei (atomic masses 12 to 56) of the calorimeter medium.

Sections \ref{sec:four} and \ref{section:nupi} below describe the analysis for
each nucleon decay mode that we studied. In each case, the particular
characteristics of the decay are reviewed and the kinematic cuts designed
to eliminate background are presented. The signal and background events
which pass the cuts are then tallied. Then, using the mode detection
efficiency and detector total exposure, a lifetime limit
$\tau /B$ at 90\% confidence level is calculated.  

\begin{table}[htb]
\begin{center}
\footnotesize
\tabcolsep 0.15cm
\footnotesize

\begin{tabular}{lccccccc|c}
 Decay & Hard- & Contain- & Event &Topology &Kinem.&BR
&$\epsilon\times$BR &$\epsilon\times$BR \\
  Mode&ware & ment & Quality &Selection & Cuts    &  & with & without \\
& Trigger & Filter & & Scans & & & INR & INR \\
\hline 
 p $\rightarrow \mu^+ \eta$, $ \eta \rightarrow \gamma \gamma $
 & 0.56 & 0.56 & 0.79 & 0.75 & 0.94 & 0.39 & 0.07$\pm$0.01 & 0.12 \\
 p $\rightarrow \mu^+ \eta$, $ \eta \rightarrow \pi^{0} \pi^{0} \pi^{0} $
 & 0.57 & 0.60 & 0.71 & 0.81 & 0.94 & 0.32 & 0.06$\pm$0.01 & 0.11 \\
 p $\rightarrow {\rm e}^+ \eta$, $ \eta \rightarrow \gamma \gamma $
 & 0.56 & 0.62 & 0.72 & 0.87 & 0.94 & 0.39 & 0.08$\pm$0.01 & 0.14 \\
 p $\rightarrow {\rm e}^+ \eta$, $ \eta \rightarrow \pi^{0} \pi^{0} \pi^{0} $
 & 0.56 & 0.62 & 0.71 & 0.88 & 0.93 & 0.32 & 0.07$\pm$0.01 & 0.12 \\
 n $\rightarrow \overline{\nu} \eta$, $ \eta \rightarrow \gamma \gamma $
 & 0.54 & 0.69 & 0.70 & 0.73 & 0.89 & 0.39 & 0.07$\pm$0.01 & 0.12 \\
 n $\rightarrow \overline{\nu} \eta$,
 $ \eta \rightarrow \pi^{0} \pi^{0} \pi^{0} $
 & 0.54 & 0.67 & 0.63 & 0.80 & 0.93 & 0.32 & 0.05$\pm$0.01 & 0.09 \\
 n $\rightarrow \overline{\nu} \pi^{0}$, $ \pi^{0} \rightarrow \gamma \gamma $
 & 0.45 & 0.73 & 0.77 & 0.57 & 0.71 & 1.0 & 0.11$\pm$0.01 & 0.21 \\
 p $\rightarrow \overline{\nu} \pi^{+}$,
 $\pi\rightarrow\mu\rightarrow{\rm e}$
 & 0.37 & 0.67 & 0.88 & 0.25 & 0.85 & 1.0 & 0.05$\pm$0.01 & 0.10 \\
\end{tabular}
\end{center}

\caption{\footnotesize Survival fractions through event selections applied
in succession, for the MC generated nucleon decay final states of this study.
The column labeled BR refers to the $\eta$ or $\pi$ decay.
The effect of intranuclear rescattering (INR) can be seen by comparing the
two rightmost columns.}

\label{tbl:efficiencies_noK}
\end{table}

\section
{Search for nucleon decay into \mbox{\boldmath $\ell \eta$} and 
\mbox{\boldmath $\overline\nu \eta$}}
\label{sec:four}

\subsection{Search for p\ \mbox{\boldmath $ \rightarrow \ell^+ \eta$}}

We have searched for proton decay into $\mu^+ \eta$ and e$^+ \eta$ and for
neutron decay into $\overline{\nu}\eta$. The decay sequences involving the
two largest branching modes of the $\eta$, namely
 $\eta \rightarrow \gamma \gamma$ and $\eta \rightarrow \pi^0 \pi^0 \pi^0$,
have been investigated. The results from both of these $\eta$ decay modes are
included in the calculation of the limits for all three nucleon decay modes.
We also explored the possibility of inclusion of\
$\eta \rightarrow \pi^+\pi^-\pi^0$ and $\eta \rightarrow \pi^+\pi^-\gamma$
into our search, but we found the resulting events
to be difficult to identify with topology criteria.
Since the potential gain is modest, we have not included these processes.

To calculate partial lifetime lower limits, $\tau/B$, we use a formalism
common to previous analyses by us \cite{Soudan_nuK+,DWall} and by the
Fr\' ejus tracking calorimeter experiment \cite{Frejus}. The mode
p $\rightarrow \mu^+ \eta$,
the first to be discussed below and which involves two daughter processes
for the $\eta$, provides an example whose generalization is straightforward:

   \begin{equation}
    (\tau / {\textstyle B})_{{\rm p} \rightarrow \mu^+ \eta^0} > N_p
 \times T_f \times 
    \frac{\left[ \epsilon_1 \times {\textstyle B}{_1}(\eta) + 
           \epsilon_2 \times {\textstyle B}{_2}(\eta) \right]}
           {\mu_1 + \mu_2}.
   \label{eq:dualLimitCalc}
   \end{equation}
Here $N_{p(n)} = 2.87 (3.15) \times 10^{32}$ protons (neutrons) in a kiloton
of the Soudan 2 detector,
$T_f = 5.52$ kiloton years is the full detector exposure,
and $\epsilon_i \times {\textstyle B}{_i}(\eta)$ are the
selection efficiencies given in Table \ref{tbl:efficiencies_noK}. The $\mu_i$
are the constrained 90\% CL upper limits on the numbers of observed events,
and are found by solving the equation
   \begin{equation}
   0.10 = \frac{\sum_{n_1 = 0}^{n_{ev;1}}\sum_{n_2 = 0}^{n_{ev;2}}
                P(n_1, b_1 + \mu_1)P(n_2, b_2 + \mu_2)}
   {\sum_{n_1 = 0}^{n_{ev;1}}\sum_{n_2 = 0}^{n_{ev;2}}P(n_1, b_1)P(n_2, b_2)}
   \label{eq:mucalc}
   \end{equation}
subject to the constraint
   \begin{equation}
    \frac{\epsilon_1 \times {\textstyle B}{_1}(\eta)}{\mu_1} = 
    \frac{\epsilon_2 \times {\textstyle B}{_2}(\eta)}{\mu_2} =
\frac{\sum_{i=1}^2 \epsilon_i \times {\textstyle B}{_i}(\eta)}{\sum_{i=1}^2 \mu_i}\ . 
   \label{eq:constraints}
   \end{equation}
In Eq. (\ref{eq:mucalc}), $P(n,\mu)$ is the Poisson function,
$e^{-\mu}\mu^n / n!$, and the $b_i$ are the estimated backgrounds.

\subsubsection*{\mbox{{\bf p}\ \boldmath $\rightarrow \mu^+ \eta$}}


\begin{figure}
\vspace{280pt}
\includegraphics{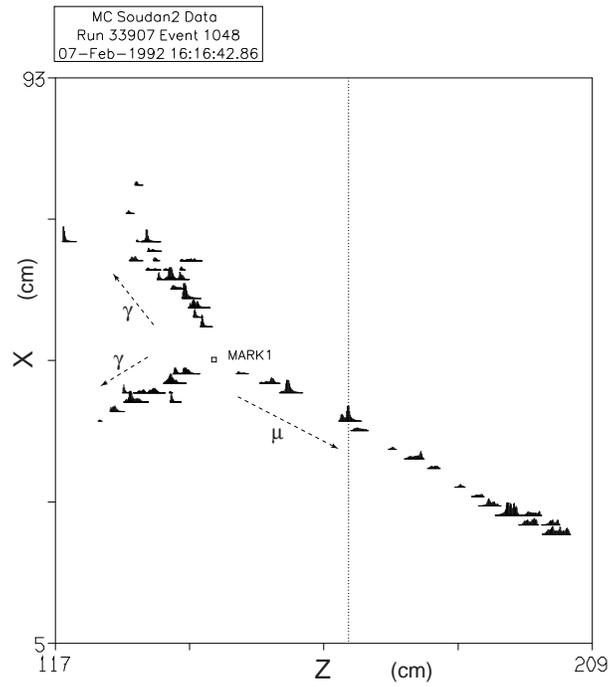}
\caption{\footnotesize Monte Carlo event with full detector response for proton
decay p $\rightarrow \mu^+\eta^0, \eta^0 \rightarrow \gamma\gamma$.
Here, the anode ($X$) versus drift time ($Z$) projection has been selected
from the three scanning views with anode-time, cathode-time, and
anode-cathode projections. The open square denoted ``MARK 1" indicates the
reconstructed primary vertex.}
\label{fig:fig1}
\end{figure}

\begin{pawepsfigfull}{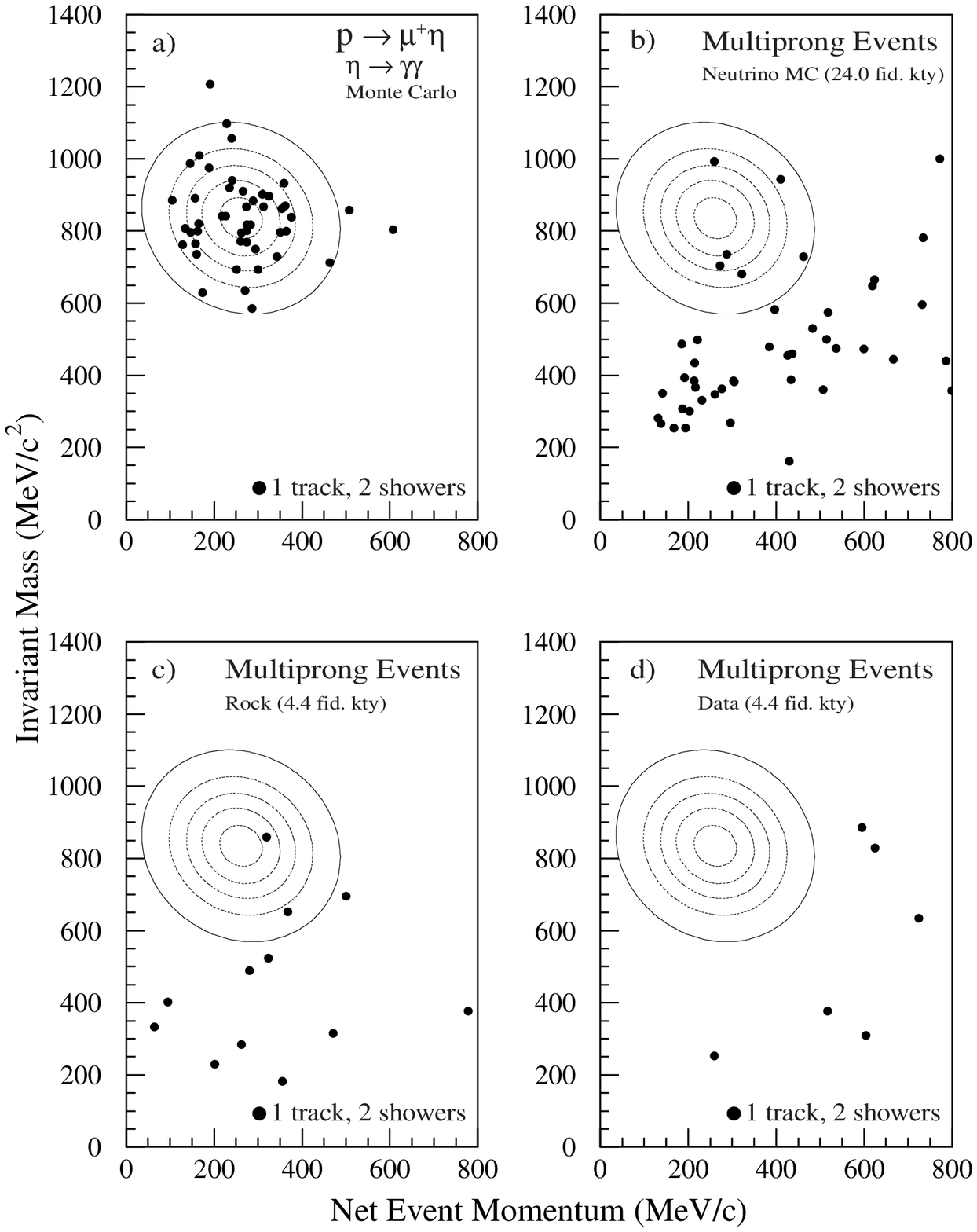}{fig:mueta_2gam}
{
{\footnotesize For p $\rightarrow \mu^+ \eta$,
$\eta \rightarrow \gamma \gamma$, event samples and the kinematic selection
contour in the $M_{inv}$ versus $\left| \vec{p}_{net} \right|$ plane.
The distributions show single track plus two shower events (solid circles)
for a) the proton decay simulation, b) the atmospheric neutrino  Monte Carlo,
c) the rock events, and d) the gold data.}
}
\end{pawepsfigfull}   

\begin{pawepsfigfull}{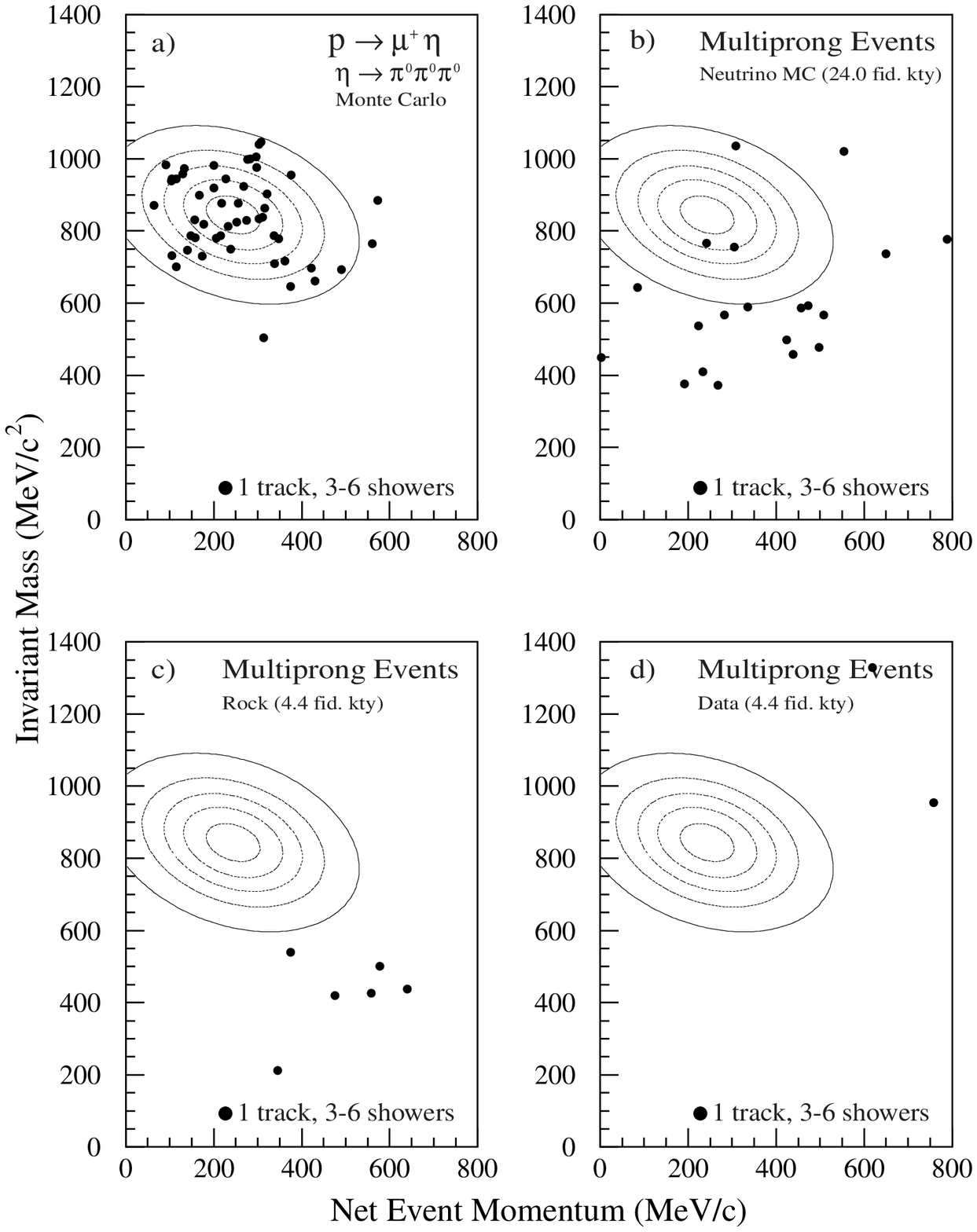}{fig:mueta_3piz}
{
{\footnotesize For p $\rightarrow \mu^+ \eta$,
$\eta \rightarrow \pi^0 \pi^0 \pi^0$, event samples and kinematic selection
contour in the $M_{inv}$ versus $\left| \vec{p}_{net} \right|$ plane.
The distributions show one track plus 3--6 shower events (solid circles) for
a) the proton decay simulation, b) the atmospheric neutrino MC,
c) the rock events, and d) the gold data events.}
}
\end{pawepsfigfull}   


      We have searched for proton decay into $\mu^+\eta$, where the $\eta$ 
decays into two photons or into three $\pi^0$ mesons.  With final states
involving either of these $\eta$ decays, the $\mu^+$ is easily
distinguished from the $\eta$ decay products in the Soudan 2
tracking calorimeter (see Fig.~\ref{fig:fig1}).
For the p $\rightarrow \mu^+\eta$  mode, the $\mu^+$ momentum is
296 MeV/$c$ in the proton rest frame.  In the laboratory frame, $\mu^+$
momenta are smeared about the two-body nominal value 
as a result of the Fermi motion of bound protons.

For p $\rightarrow \mu^+\eta$, $\eta \rightarrow \gamma\gamma$,
the kinematic region in the invariant mass versus net momentum
plane which contains 90\% of the reconstructed MC proton decay events,
is delineated by the outer contour displayed in
Fig.~\ref{fig:mueta_2gam}. Since the entire final state is
visible, the invariant mass will approximate the nucleon mass and the
net event momentum will be distributed in accordance with the convolution
of Fermi motion with detector resolution. Fig. \ref{fig:mueta_2gam} shows the
distribution of events within this plane, for
the proton decay simulation sample, for the atmospheric neutrino MC
sample, and for the rock and gold data samples.  The total background
which arises from neutrino interactions and from the cosmic ray induced
rock events is 0.9 + 0.1 = 1.0 events.   The neutrino background originates from
assorted multiple pion production channels which, after correction for
$\nu_\mu$ oscillations, is comprised of equal portions of $\nu_\mu$ and
$\nu_e$ inelastic charged-current events.

No events are observed to pass the primary kinematic contour. Due to the small
background expectation, no additional kinematic constraints (e.g. on the
momentum of the $\eta$ or the $\mu^+$) were applied. With an overall
detection efficiency of 6.9\% we establish a limit of $\tau / B > 48 \times
10^{30}$~years at 90\% CL for this decay sequence.

In the other decay sequence of p $\rightarrow \mu^+ \eta$, namely
$\eta \rightarrow \pi^0 \pi^0 \pi^0$, we search for a decay signature of one
track and three to six decay showers from the
three $\pi^0$s. The vertex is defined as the end of the track from which the
showers emerge. In this case, the rate of the rock background is small and
the background expectation of 0.5 events arises almost entirely from
atmospheric neutrino interactions. Fig.~\ref{fig:mueta_3piz} shows the kinematic
search regions in the $M_{inv}$ versus $\left| \vec{p}_{net} \right|$ plane,
together  with the event distributions from the four samples.
The detection efficiency for this decay sequence
is 5.9\%. No data events are observed to pass the kinematic cuts.
The limit for this decay sequence is calculated to be
$\tau / B > 41 \times 10^{30}$~years at 90\% CL.
Combining the two daughter processes we obtain an overall limit of
$\tau / B > 89 \times 10^{30}$~years at 90\% CL.

\subsubsection*{\mbox{{\bf p} \boldmath $\rightarrow {\rm e}^+ \eta$}}

\begin{pawepsfigfull}{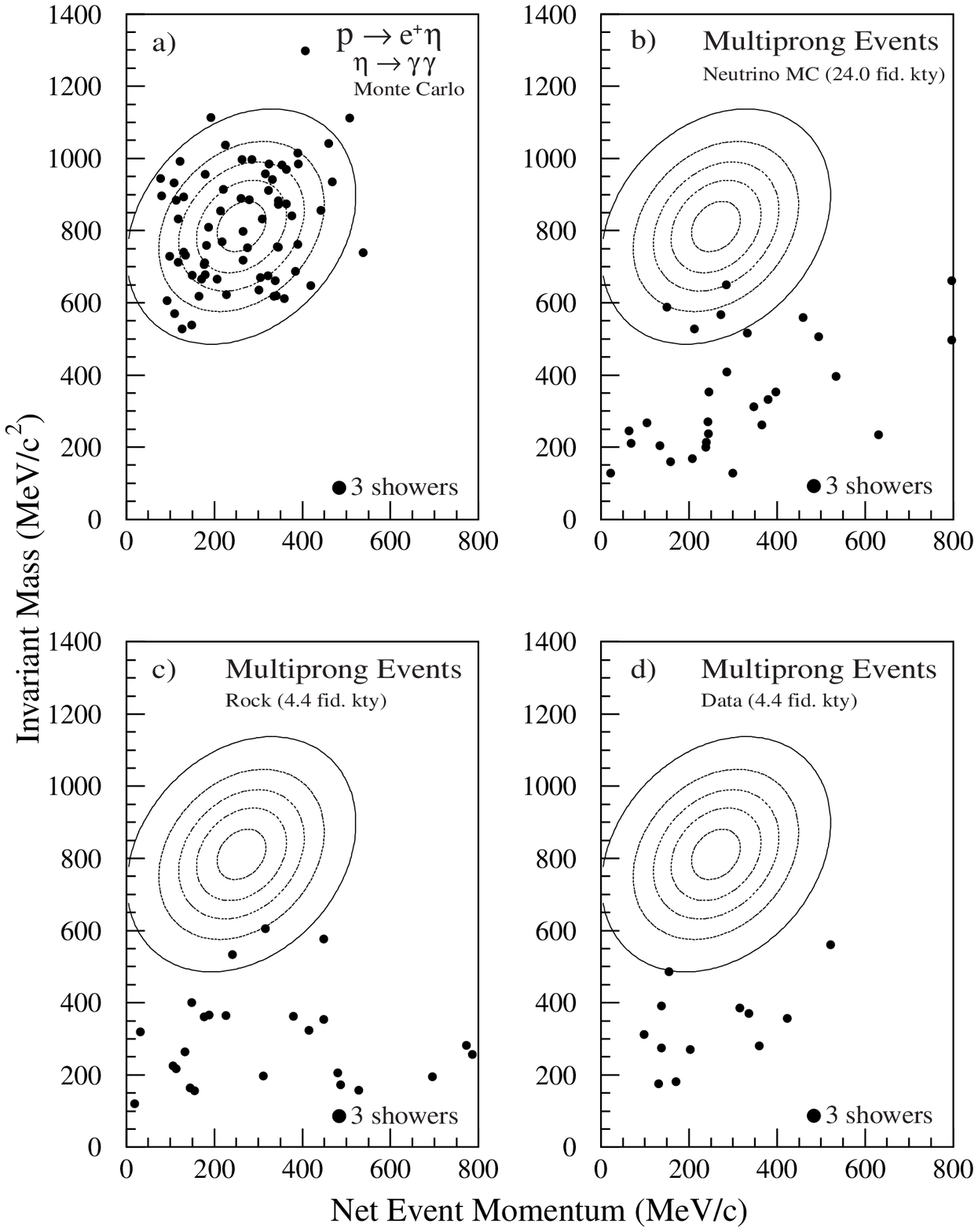}{fig:eeta_2gam}
{
{\footnotesize For p $\rightarrow e^+ \eta$, $\eta \rightarrow \gamma \gamma$,
event samples and kinematic selection contour
in the $M_{inv}$ versus $\left| \vec{p}_{net} \right|$ plane. Distributions
show a) the proton decay simulation, b) three-shower events of the atmospheric
neutrino MC, c) three-shower rock events, and d) three-shower events of gold
data.}
}
\end{pawepsfigfull}  

\begin{pawepsfigfull}{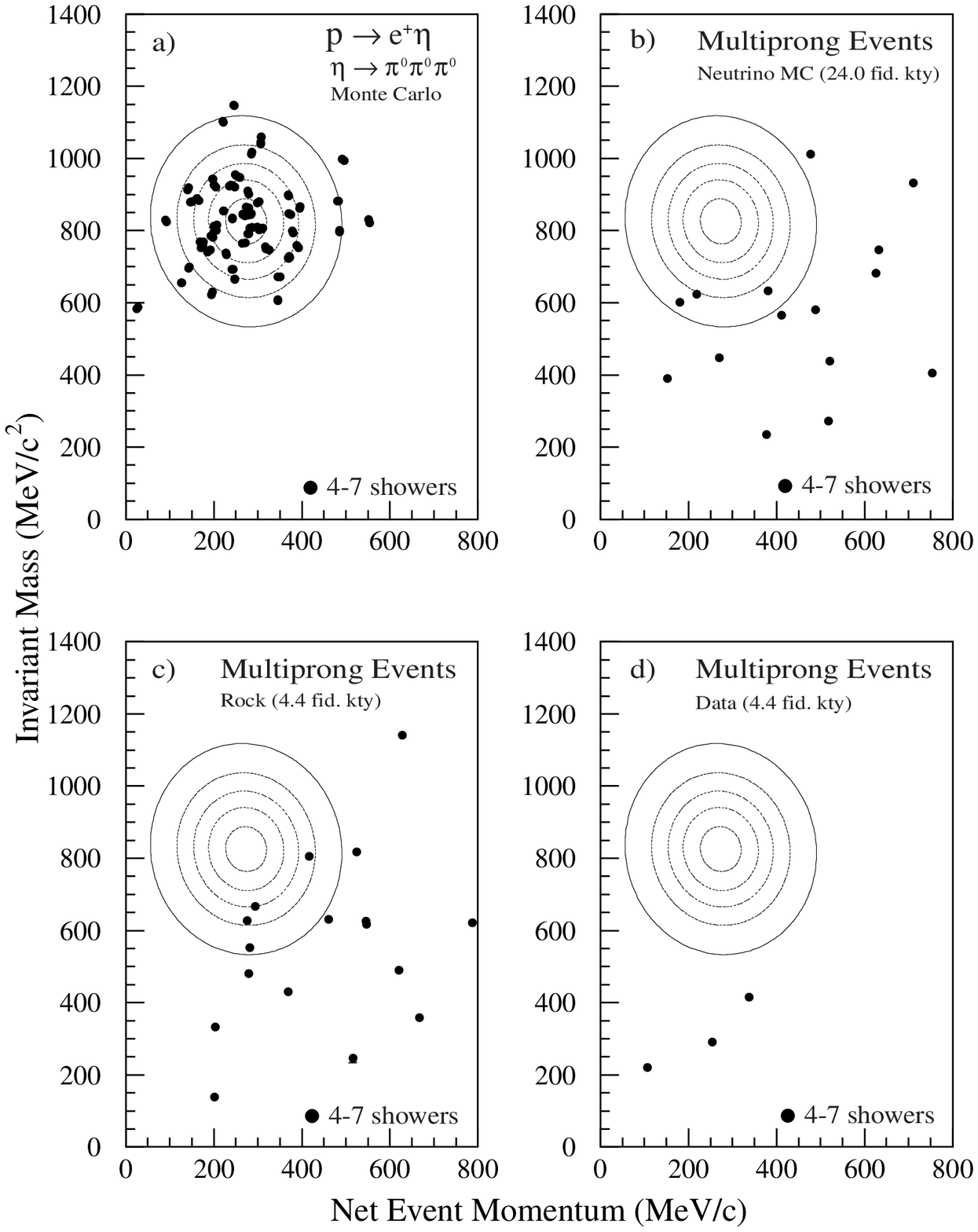}{fig:eeta_3piz}
{
{\footnotesize For p $\rightarrow e^+ \eta$,
 $\eta \rightarrow \pi^0 \pi^0 \pi^0$, event samples and kinematic selection
contour in the $M_{inv}$ versus $\left| \vec{p}_{net} \right|$ plane.
Distributions show a) the proton decay simulation, b) 4--7 shower events of
the atmospheric neutrino MC, c) 4--7 shower rock events, and d) 4--7 shower
events of gold data.}
}
\end{pawepsfigfull}

For proton decay into e$^+ \eta$ with $\eta \rightarrow \gamma \gamma$, we
searched for an event topology of three distinct showers emanating from a
common vertex. We found it was not advantageous to distinguish the primary
positron shower from the decay photons of the $\eta$; rather, an overall
search in the $M_{inv}$ versus $\left| \vec{p}_{net} \right|$ plane was
conducted. Fig.~\ref{fig:eeta_2gam} shows all three-shower events and their
relation to the kinematic cut region for the proton decay MC, atmospheric
neutrino Monte Carlo, rock, and gold data samples. The total
background estimate is 0.9 events, of which 0.7 events are calculated to be
initiated by atmospheric neutrinos. The background events are mostly due to
$\nu_e$ and $\overline{\nu}_e$ inelastic charged current interactions. The
number of nucleon decay candidates observed is one event. From this we deduce
a partial lifetime limit of $\tau / B > 38 \times 10^{30}$~years at 90\% CL. 

For the $\eta \rightarrow \pi^0 \pi^0 \pi^0$ mode, the signal topology is
taken to be four to seven showers, all emerging from a common vertex.
The detection efficiency is calculated to be 6.5\%. As with
$\eta \rightarrow \gamma \gamma$, no attempt is made to identify which of the
several showers in the event was from the prompt positron. Distributions for
this mode are depicted in Fig.~\ref{fig:eeta_3piz}. Most of the background
expectation of 0.6 events can be attributed to $\nu_e$ charged current
multiple $\pi^0$ production. No data events were observed to pass the
kinematic cuts for this mode, and we obtain a limit of
$\tau / B > 48 \times 10^{30}$~years at 90\% CL. Combining the two
$\eta$ decay modes gives an overall limit for p $\rightarrow$ e$^+\eta$
of $\tau / B > 81 \times 10^{30}$~years at 90\% CL.

\subsubsection*{\mbox{{\bf n} \boldmath $ \rightarrow \overline{\nu}\eta$}}

\begin{pawepsfigfull}{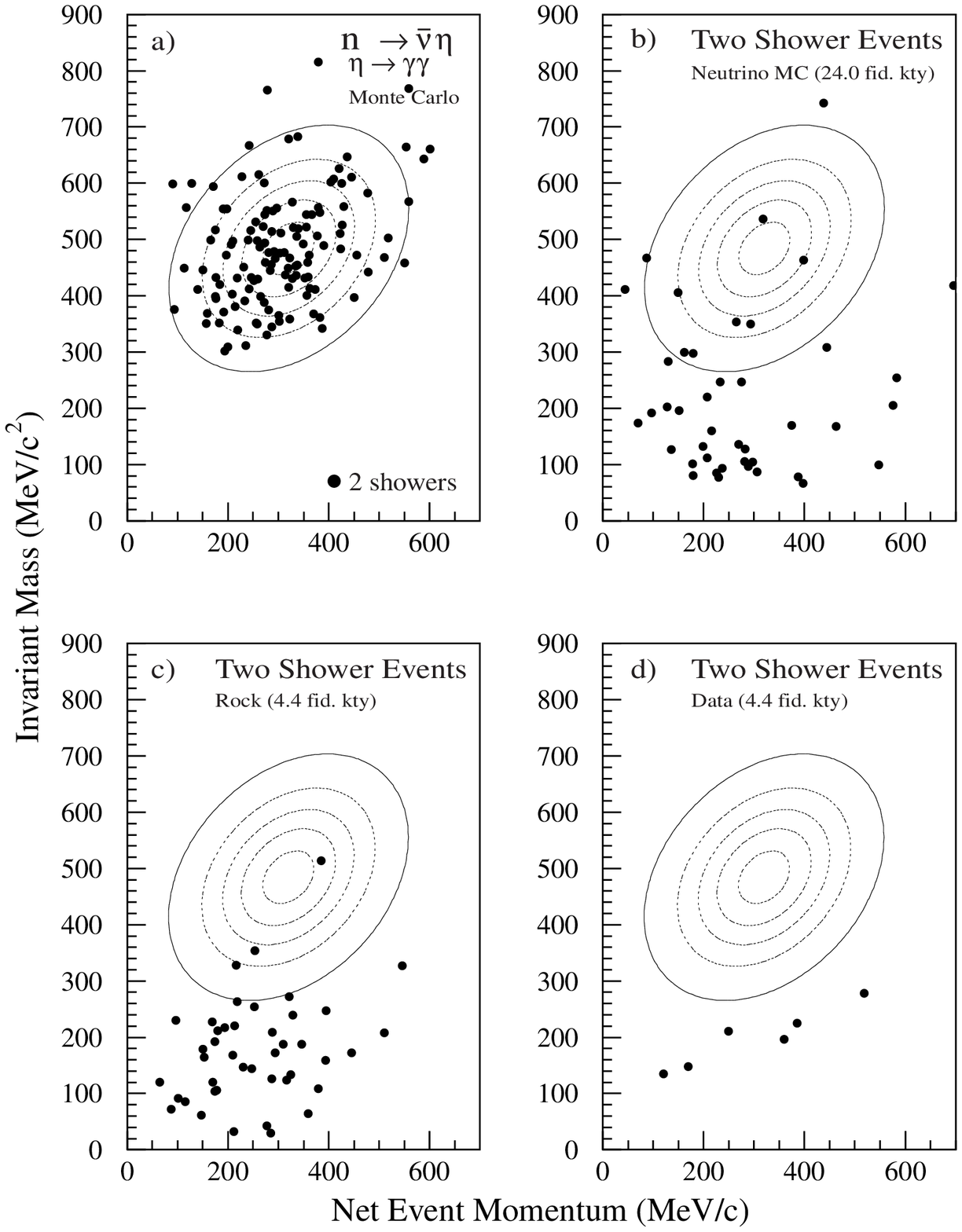}{fig:nueta_1}
{
{\footnotesize For n $\rightarrow \overline{\nu} \eta$,
$\eta \rightarrow \gamma \gamma$, event samples and kinematic selection contour
in the $\gamma\gamma$ $M_{inv}$ versus $\left| \vec{p}_{net} \right|$ plane.
Distributions show a) the neutron decay simulation, b) two-shower events
of the atmospheric neutrino MC, c) two-shower events of the shield-tagged
``rock" sample, and d) two-shower events of gold data.}
}
\end{pawepsfigfull}

\begin{pawepsfigfull}{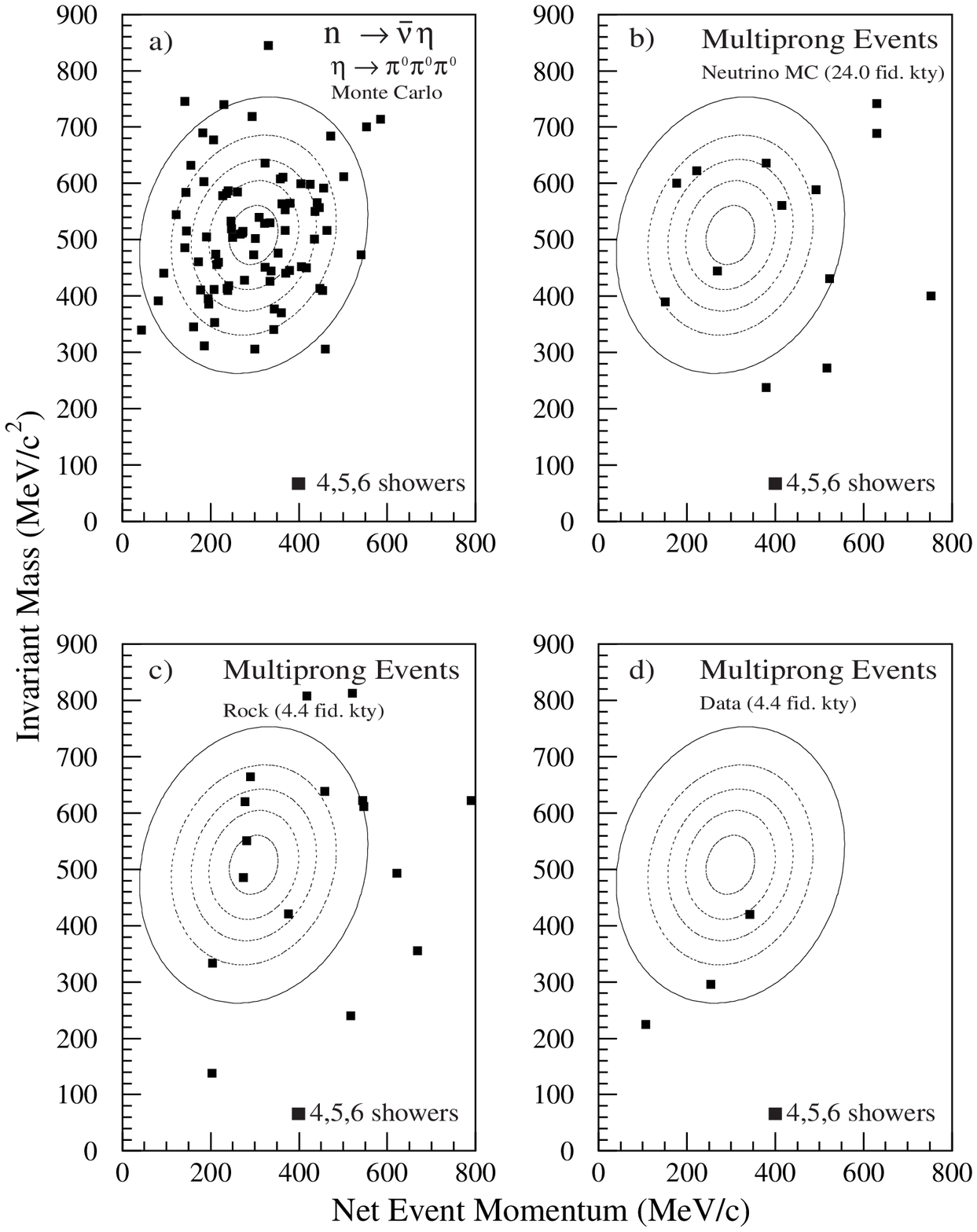}{fig:nueta_3pz}
{
{\footnotesize For n $\rightarrow \overline\nu \eta$,
 $\eta \rightarrow \pi^0 \pi^0 \pi^0$, event samples and kinematic selection
contour in the $M_{inv}$ versus $\left| \vec{p}_{net} \right|$ plane.
Distributions show a) the nucleon decay simulation, b) 4-6 shower events of
the atmospheric neutrino MC, c) 4--6 shower rock events, and d) 4--6 shower
events of gold data.}
}
\end{pawepsfigfull}

Neutron decay n $\rightarrow \overline{\nu}\eta$ with
$\eta \rightarrow \gamma \gamma$  involves a two--shower final state
originating from a ``sharp'' invariant mass of 547 MeV/$c^2$. We define for
n $\rightarrow \overline{\nu} \eta$, $\eta \rightarrow \gamma\gamma$,
a kinematically--allowed elliptical region in the
$m_{\gamma\gamma}$ versus $\left| \vec{p}_{net} \right|$ plane, as shown
in Fig.~\ref{fig:nueta_1}.
The four diplots of Fig.~\ref{fig:nueta_1} show the effect of the kinematic
contour selection on the neutron decay simulation (Fig.~\ref{fig:nueta_1}a)
and on the two-shower events from the atmospheric neutrino MC, from the
rock event sample, and from the gold data. The kinematic contour selection
is satisfied by 90\% of the MC $\overline{\nu} \eta$ events,
whereas the neutrino and rock backgrounds are almost entirely eliminated.
With the loss due to intranuclear rescattering, our detection efficiency
for n $\rightarrow \overline{\nu} \eta$, $\eta \rightarrow \gamma\gamma$
becomes 7\%. For this mode, there are no candidate events. The background from
neutrino and from rock events is estimated to be 1.7 events.
Nearly half of the neutrino-induced background arises from neutral-current
production of $\eta$ and $\pi^0$'s. A significant contribution
also arises from $\nu_e$ charged current single $\pi^0$ production events in
which one of the showers is not discernible in scanning. For
n $\rightarrow \overline{\nu} \eta_{\gamma\gamma}$ we obtain a partial
lifetime lower limit of $\tau / B > 53 \times 10^{30}$~years at 90\% CL.

For the other $\eta$ decay mode considered,
$\eta \rightarrow \pi^0 \pi^0 \pi^0$, the signal topology is four to six
showers emerging from a common vertex. The detection efficiency for
n $\rightarrow \overline{\nu} \eta_{\pi^0\pi^0\pi^0}$ is 5.4\% including the
intranuclear correction. The distributions in the plane of $M_{inv}$
versus $\left| \vec{p}_{net} \right|$ are depicted in Fig.~\ref{fig:nueta_3pz}.
From Fig.~\ref{fig:nueta_3pz}b,
 the total neutrino background is estimated to be 1.5 events,
of which nearly half is $\nu_e$ charged current interactions with multiple
pion production. The remaining background is due to inelastic
neutral current interactions. Another 0.6 events are expected from rock
events, bringing the total background expectation to 2.0 events. We observe
two data events and this gives a partial lifetime limit of $\tau / B > 22 \times
10^{30}$~years at 90\% CL. Combining the two submodes together yields a limit
for n $\rightarrow \overline{\nu}\eta$
of $\tau / B > 71 \times 10^{30}$~years at 90\% CL.

\section{SEARCH FOR
 \mbox{\boldmath ${\rm n}\ \rightarrow \overline{\nu} \pi^0$}
 AND \mbox{\boldmath ${\rm p}\ \rightarrow \overline{\nu} \pi^+$}}

\label{section:nupi}

We have searched for the two-body nucleon decay modes which yield
a final-state

\noindent (anti-)neutrino together with a $\pi$--meson.
For the neutron decay into $\overline{\nu} \pi^0$, the observable
final state consists of two photon showers having a restricted invariant mass,
and discrimination from background can be carried out similarly to
n $\rightarrow \overline{\nu} \eta$, $\eta \rightarrow \gamma\gamma$.

For proton decay into $\overline{\nu} \pi^+$, the observable final state is
simply a single charged track which can scatter and/or range to stopping with
an endpoint decay ($\pi^+ \rightarrow \mu^+ \rightarrow {\rm e}^+$).
A restricted allowed range of pion momentum is implied by the two--body nature
of this proton decay. However, the background from quasi--elastic $\nu_{\mu}$
and ${\overline{\nu}}_{\mu}$ neutrinos with unobserved proton and neutron
recoils is substantial, since muon tracks are indistinguishable from $\pi^+$
tracks in the Soudan calorimeter, unless of course a distinct scatter is
present. Consequently, for the p $\rightarrow \overline{\nu} \pi^+$ search
reported below, we require that an endpoint decay be present on candidate
single track events. This requirement discriminates against the background from
quasi--elastic $\nu_{\mu} {\rm n} \rightarrow \mu^- {\rm p}$ reactions
(though ${\overline{\nu}}_{\mu} {\rm p} \rightarrow \mu^+ {\rm n}$ background
still remains).

\subsubsection*{\mbox{{\bf n} \boldmath $ \rightarrow \overline{\nu} \pi^0$}}

\begin{pawepsfigfull}{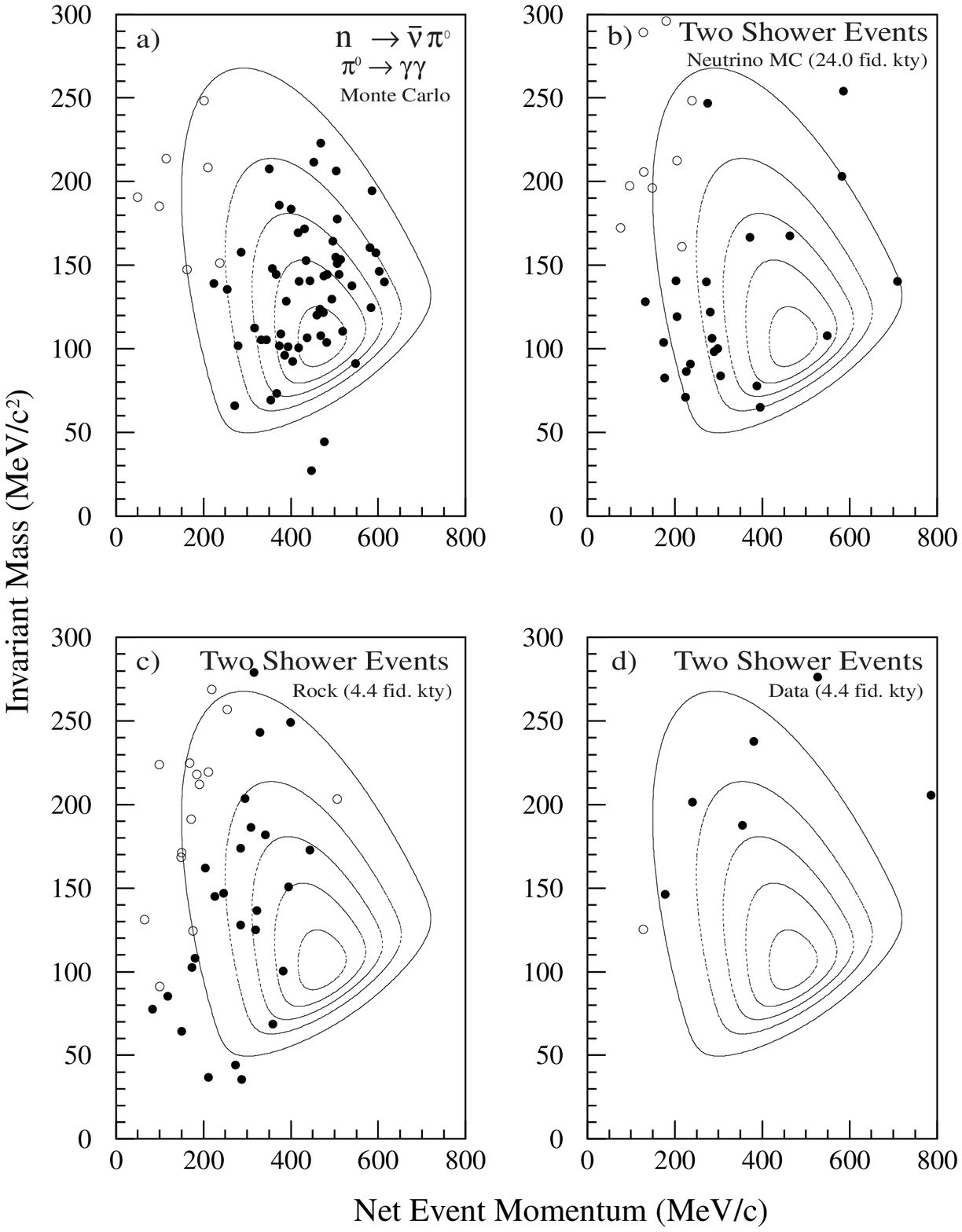}{fig:nupi0_1}
{
{\footnotesize
For n $\rightarrow \overline{\nu} \pi^0$, event samples and kinematic
selection contour in the $M_{inv}$ versus $\left| \vec{p}_{net} \right|$ plane.
Distributions show two shower events (a) of the nucleon decay simulation,
(b) of the atmospheric neutrino MC, (c) of the shield-tagged rock events,
and (d) of gold data. Events depicted with solid circles satisfy an
angular separation requirement on the gamma showers.}
}
\end{pawepsfigfull}   

Neutron decay into $\overline{\nu} \pi^0$ yields a two--shower final state.
The invariant mass values obtained by reconstructing event
shower pairs from the nucleon decay MC simulation comprise the $y$-coordinates
of the events shown in Fig.~\ref{fig:nupi0_1}a. The $\gamma\gamma$ mass
distribution peaks close to the $\pi^0$ mass, however it is rather broad
(a mean of 142 MeV/$c^2$ with a sigma of 45 MeV/$c^2$ from a Gaussian fit).
For $\overline{\nu} \pi^0$, a bi--variate Gaussian does not provide
a good characterization of the event distribution in the $M_{inv}$ versus
$\left| \vec{p}_{net} \right|$ plane.
For the purpose of accommodating the data, a two--dimensional Gaussian was
multiplied by a sigmoid function (a smoothed step function).
In fitting the data, parameters governing the orientation of the sigmoid and the slope of its step were
allowed to vary, in addition to the parameters of the Gaussian. The result of
this fit is the set of contours shown in Fig.~\ref{fig:nupi0_1}.
As an additional constraint, we require that the two showers
have an opening angle less than $90^{\circ}$. Events which are thereby
eliminated are shown as open circles in Fig.~\ref{fig:nupi0_1}.

In our  n $\rightarrow \overline{\nu} \pi^0$ simulation, 86\% of events lie
within the outer contour of Fig.~\ref{fig:nupi0_1}a and also satisfy the 
$\gamma\gamma$ opening angle requirement. The overall detection efficiency
for this mode is 11\%. For a simulation in which the intranuclear effects are
not included we find a detection efficiency of 21\%. The bulk of the INR losses
can be attributed to intranuclear absorption and inelastic scattering
processes which lower the trigger rate for the simulation from 87\% to 45\%.
A similar INR effect is observed for p $\rightarrow \overline{\nu} \pi^+$.

For n $\rightarrow \overline{\nu} \pi^0$ we calculate that 3.8 background
events are expected, with 0.9 events from rock processes and 2.9 events
from neutrino background. Inspection of the neutrino background reveals that
75\% of the events are neutral-current inelastic single $\pi^0$ production
events in which the recoil baryon is either a neutron that escapes detection
or a proton that is produced below threshold. The remaining 25\% of the
background is due to $\nu_e$ charged-current single charged pion production
events in which the pion is misidentified as a shower and recoil baryons
are undetected. In the gold data we observe 4 candidate events. The
background--subtracted lifetime lower limit at 90\% CL is then
$\tau / B > 39 \times 10^{30}$~years.

\subsubsection*{\mbox{{\bf p} \boldmath $ \rightarrow \overline{\nu} \pi^+$}}

   
\begin{pawepsfigfull}{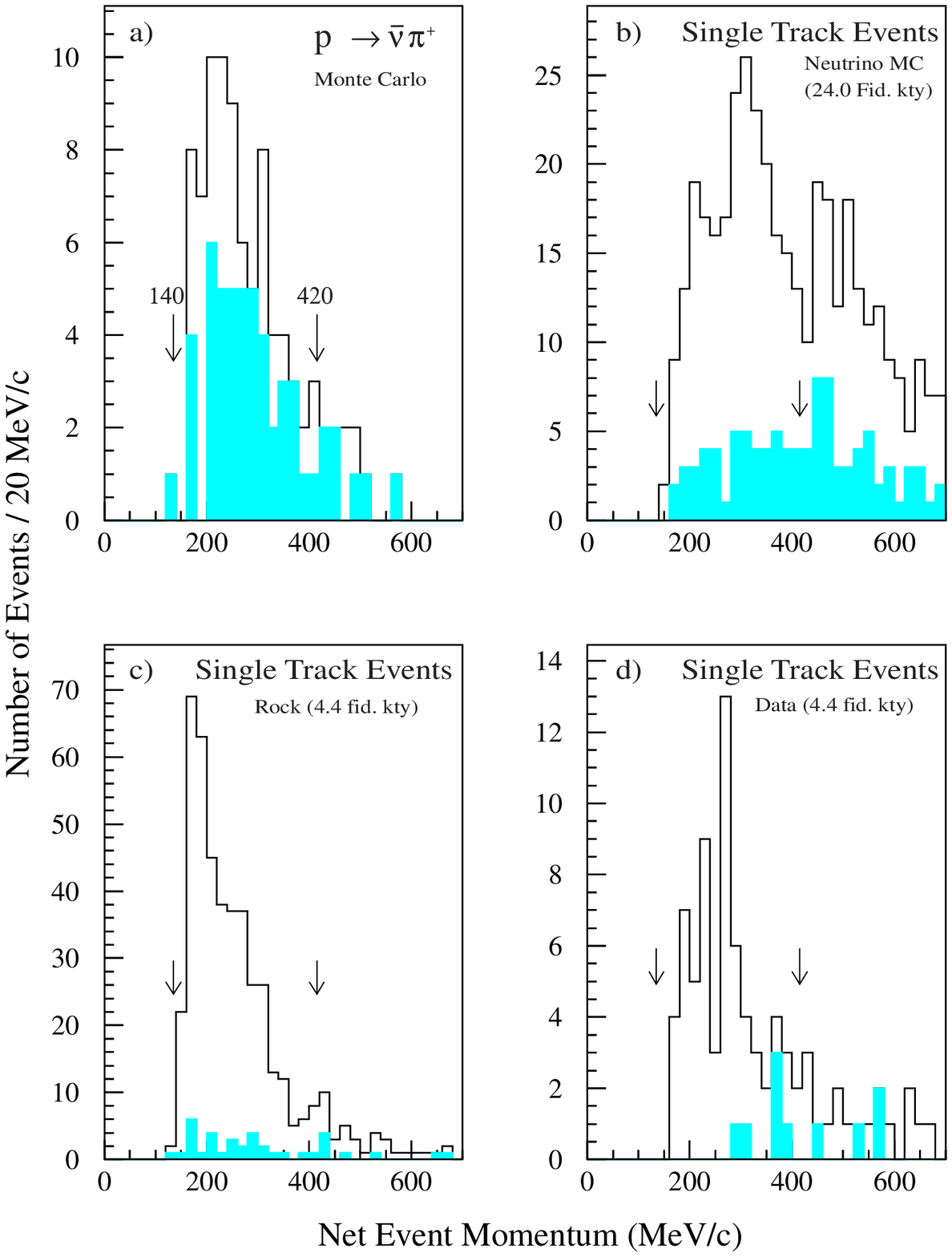}{fig:pnupi_plus_quad}
{
{\footnotesize For p $\rightarrow \overline{\nu} \pi^+$, the track
momentum distribution from a) the proton decay simulation, b) single track
events of the atmospheric neutrino MC, c) single track rock events,
and d) single track events of gold data. Events having visible endpoint
decays on candidate tracks are shown with shaded histogramming.}
}
\end{pawepsfigfull}   

The two--body decay p $\rightarrow \overline{\nu} \pi^+$, 
of an unbound stationary proton would produce a $\pi^+$ with a momentum of
459 MeV/$c$. In the Soudan 2 calorimeter, however, protons are mostly to be
found within iron nuclei. Pions which are created inside such a nucleus can
undergo intranuclear rescattering before they emerge; and, even if they emerge
with their identity intact, the $\pi^+$ mesons traverse a dense medium
where they can undergo (further) large energy degradation due to scattering
processes. The net
result is that $\pi^+$ mesons of this decay mode exhibit only half
of their initial momentum on average, as can be seen from the result of our full
simulation shown in Fig.~\ref{fig:pnupi_plus_quad}a. For our search we require
candidate events to have one and only one track (no recoil proton or neutron)
with ionization compatible with a pion or muon mass assignment. The pion
momentum as reconstructed from range is required to fall within
140 to 420 MeV/$c$. Additionally, the track is required to have a visible
endpoint decay consisting of two or more decay shower hits. These cuts reduce
the final-state detection efficiency by roughly 50\% but remove enough of the
neutrino background rate to make a search feasible. The total detection
efficiency for this mode is 4.6\%.

To gauge the effects on efficiency of intranuclear rescattering
within the parent nuclei of the calorimeter
medium, a proton decay simulation that does not include
intranuclear effects has been compared with the full
simulation. We observe that the trigger efficiency increases from 47\% to 72\%
in the absence of intranuclear effects. Additionally, the average momentum of
the reconstructed pions of the simulation increases from 284 MeV/$c$ to
356 MeV/$c$. Evidently, part of the
discrepancy between the predicted two body decay momentum of 459 MeV/$c$ and
the average momentum of the reconstructed pion tracks of the simulation can be
attributed to intranuclear scattering. Hadronic scattering processes in the
detector medium account for the remaining difference. Finally,
in the absence of intranuclear effects, the overall efficiency for
p $\rightarrow \overline{\nu} \pi^+$ would be increased from 4.6\% to 8.8\%. 

With the cuts optimized as described above, we observe 6 candidate events and
we estimate the background to be 10.5 events in the absence of oscillations
by atmospheric neutrinos.
However, in the presence of atmospheric $\nu_\mu \rightarrow \nu_\tau$
oscillations, our background estimate must be scaled by the Soudan-2 flavor 
ratio (see Section IIC) and is thereby reduced to 7.7 events.
The lifetime lower limit at 90\% CL is then 
$\tau / B > 16 \times 10^{30}$~years.

\section{Uncertainties and lifetime limits}
         
     The lifetime lower limits reported here are affected by
statistical and systematic uncertainties which arise with detection of
each nucleon decay final state and with background estimation.  The various 
error sources, and the corresponding fractional variation 
$\Delta\tau_N/\tau_N$ thereby introduced, are similar to those detailed
for our lepton + K$^0$ modes search in Soudan 2
(see Ref.~\cite{DWall}, Sect. V). The one exception lies with treatment of
intranuclear rescattering losses within parent nuclei
for the $\eta$ and $\pi$ modes studied here.
  In contrast to produced K$^0$s (strangeness = +1),
eta mesons and pions may have sizeable rescattering probability for which there
is also significant uncertainty.
  Based upon uncertainties which arise in
our phenomenological cascade model \cite{mann:intranuke:leeson:pdk678},
with extrapolation to heavier nuclei of
pion production observed in $\nu_\mu$ deuteron ($A=2$) and $\nu_\mu$ neon
($A=20$) reactions \cite{merenyi}, we estimate an uncertainty of 30\% for our 
rescattering treatment. This error is to be added in quadrature to the 
errors (see below) listed for individual channel detection efficiencies,
$\epsilon \times BR$, in Table I.
 As can be seen from equation (\ref{eq:dualLimitCalc}),
the INR uncertainty enters directly into $\Delta\tau_N/\tau_N$ via the
detection efficiencies $\epsilon_i$.

     For each channel there is accumulated error on the survival
efficiency through selections imposed by triggering, software filtering,
scanning, and kinematic cuts; this can be as large as 18\%, as indicated by
the next-to-rightmost column in Table~\ref{tbl:efficiencies_noK}. 
For the $\overline{\nu} \pi$ modes, and also
for $\overline{\nu} \eta$, errors enter the lifetime limit through the estimates
of background from atmospheric neutrino events and from cosmic ray induced
rock events.   Propagation of background errors through relations
(\ref{eq:dualLimitCalc}), (\ref{eq:mucalc}), and (\ref{eq:constraints})
for individual channels gives $\Delta\tau_N/\tau_N \le 20\%$.
We conclude that the uncertainty $\Delta\tau_N/\tau_N$  on the lifetime
lower limits reported in this work may be as large as 40\%.
Of course, comparable uncertainties also apply to other published limits
on the nucleon lifetime.

\section{Summary and conclusions}

A search for five distinct lepton + $\eta (\pi)$
nucleon decay channels has been carried out using a
fiducial exposure of 4.4 kiloton--years recorded by the Soudan 2 fine-grained
iron tracking calorimeter. The modes considered are among those
proposed in supersymmetric grand unified models. For each mode,
cuts have been designed which minimize cosmic-ray neutrino--, photon-- and
hadron--induced background while maintaining sufficient detection
efficiency to allow a sensitive search. From among all of the
lepton + pseudoscalar meson modes investigated, zero or small numbers of
candidate events are observed; in every case the occurrence of candidates is
compatible with expectations for background. A summary of our
partial lifetime lower limits $\tau / B$ at 90\% CL obtained with each channel
is given in Table~\ref{tbl:summary}.

\begin{table}[htb]

\tabcolsep 0.15cm
\footnotesize

\begin{tabular}{lcccccccc}
Decay Mode&Final State&$\epsilon\times$B.R.&
 \multicolumn{3}{c}{Background}&Data&N$_{90}$&$\tau /B \times 10^{30}y$ \\
 \cline{4-6}
& & & $\nu$&Rock&Total\\
\hline
p$\rightarrow \mu^+ \eta $ & $\gamma \gamma $        & 0.07 & 0.9(1.1) & 0.1 
& 1.0  & 0  & 2.3 & 89 \\
p$\rightarrow \mu^+ \eta $ & $\pi^0\pi^0\pi^0$ & 0.06 & 0.5(0.6) &$<0.06$
& 0.6 & 0  &     &     \\
p$\rightarrow e^+ \eta $   & $\gamma \gamma $        & 0.08 & 0.7 & 0.1 
& 0.9 & 1  & 2.9 & 81  \\
p$\rightarrow e^+ \eta $   & $\pi^0\pi^0\pi^0$ & 0.07 & 0.6 & 0.2 
& 0.8 & 0  &     &     \\
n$\rightarrow \overline{\nu} \eta $     & $\gamma \gamma $        & 0.07
& 1.5 & 0.2 & 1.7 & 0  & 2.9 & 71  \\
n$\rightarrow \overline{\nu} \eta $     & $\pi^0\pi^0\pi^0$ & 0.05 & 1.5
& 0.6 & 2.0 & 2  &     &     \\
n$\rightarrow \overline{\nu} \pi ^0$    & $\gamma \gamma $        & 0.11
& 2.9  & 0.9 & 3.8  & 4  & 4.8 & 39  \\
p$\rightarrow \overline{\nu} \pi ^+$  & $\pi^+$               & 0.05 & 5.0(8.8)
& 1.7  & 7.7 & 6  & 4.0 & 16 \\
\end{tabular}

\caption{\footnotesize Background--subtracted lifetime lower limits at 90\%
confidence level from Soudan 2. Atmospheric $\nu_\mu$ depletion due to
neutrino oscillations has discernible effect for neutrino background to
p $\rightarrow \overline{\nu}\pi^+$, and also to
p $\rightarrow \mu^+\eta$; background rates for null oscillations are
shown in parentheses.}
\label{tbl:summary}
\end{table}

A comparison of our current Soudan 2 limits with results published by the
Fr\'{e}jus, Kamiokande, and IMB-3 experiments is presented in
Table~\ref{tbl:complimitsnew}.
 For the two-body lepton-plus-eta decay modes
p $\rightarrow \mu^+ \eta$, p $\rightarrow {\rm e}^+ \eta$, and
n $\rightarrow \overline{\nu} \eta$, we observe zero, one, and two candidates
respectively, with comparable expectations for background processes.
Our resulting lifetime lower limits are compatible with those listed by the
Particle Data Group \cite{pdg:rpp} and are the most stringent limits achieved
using iron calorimeters.

\begin{table}
\begin{center}
\footnotesize

\begin{tabular}{lccccc}
& & \multicolumn{4}{c}{$\tau /B\ (10^{30}$ years)}\\ \cline{3-6}
Decay Mode & & Soudan 2 & Fr\'{e}jus \cite{Frejus,Frejus2} 
& Kamiokande \cite{Kamiokande} & IMB--3 \cite{IMB-3} \\
\hline 
p $\rightarrow \mu^+ \eta$& &  89 & 26 & 69 & 126  \\
p $\rightarrow {\rm e}^+ \eta$& & 81 & 44 & 140 & 313  \\
n $\rightarrow \overline{\nu} \eta$& & 71 & 29 & 54 & 158  \\
n $\rightarrow \overline{\nu} \pi^0$& & 39 & 13 & 100 & 112  \\
p $\rightarrow \overline{\nu} \pi^+$& & 20 & 10 & 25 & 10  \\
\end{tabular}
\end{center}

\caption{\footnotesize Comparison of 90\% confidence level
 background-subtracted limits in nucleon decay experiments.}

\label{tbl:complimitsnew}
\end{table}

\eject

\section*{Acknowledgements}

This work was supported by the U.S. Department of Energy, the U.K. Particle
 Physics and Astronomy Research Council, and the State and University of
 Minnesota. We wish to thank the Minnesota Department of Natural
 Resources for allowing us to use the facilities of the Soudan Underground
Mine State Park.


\end{document}

%% file: ltkmcrs.tex
\def\beq{\begin{eqnarray}}
\def\beqs{\begin{eqnarray*}}
\def\eeq{\end{eqnarray}}
\def\eeqs{\end{eqnarray*}}

\newcommand{\fecha}{\ifcase\month\or
January\or February\or March\or April\or May\or June\or
July\or August\or September\or October\or November\or December\fi
\space\number\day, \number\year}

\newcommand{\lastlabel}{lastlabel}

{
\begin{figure}
     \vspace{3.5in} 
     \includegraphics{#1}
	\renewcommand{\lastlabel}{#2}
	\caption
}
{\label{\lastlabel}\end{figure}}

{
\begin{figure}[p]
     \vspace{7.0in} 
     \includegraphics{#1}
	\renewcommand{\lastlabel}{#2}
	\caption
}
{\label{\lastlabel}\end{figure}}

{
\begin{figure}
     \vspace{3.5in} 
     \includegraphics{#1}
	\renewcommand{\lastlabel}{#2}
	\caption
}
{\label{\lastlabel}\end{figure}}

\newenvironment{pawepsfigfull}[2]%
{
\begin{figure}[p]
     \vspace{7.0in} 
     \includegraphics{#1}
	\renewcommand{\lastlabel}{#2}
	\caption
}
{\label{\lastlabel}\end{figure}}

{
\begin{figure}[p]
     \vspace{3.5in} 
     \includegraphics{#1} 
	\renewcommand{\lastlabel}{#2}
	\caption
}
{\label{\lastlabel}\end{figure}}

{
\begin{figure}[p]
     \vspace{3.5in} 
     \includegraphics{#1}
	\renewcommand{\lastlabel}{#2}
	\caption
}
{\label{\lastlabel}\end{figure}}

{
\begin{figure}
     \vspace{3.5in} 
     \includegraphics{#1}
	\renewcommand{\lastlabel}{#2}
	\caption
}
{\label{\lastlabel}\end{figure}}

{
\begin{figure}
     \vspace{7.0in} 
     \includegraphics{#1}
	\renewcommand{\lastlabel}{#2}
	\caption
}
{\label{\lastlabel}\end{figure}}

{
\begin{figure}
     \vspace{7.0in} 
     \includegraphics{#1}
	\renewcommand{\lastlabel}{#2}
	\caption
}
{\label{\lastlabel}\end{figure}}

{
\begin{figure}
     \vspace{3.5in} 
     \includegraphics{#1} 
	\renewcommand{\lastlabel}{#2}
	\caption
}%
{\label{\lastlabel}\end{figure}}

{
\begin{figure}[h]
     \vspace{3.5in} 
     \includegraphics{#1} 
	\renewcommand{\lastlabel}{#2}
	\caption
}%
{\label{\lastlabel}\end{figure}}

{
\begin{figure}[p]
     \vspace{3.5in} 
     \includegraphics{#1} 
	\renewcommand{\lastlabel}{#2}
	\caption
}%
{\label{\lastlabel}\end{figure}} 

{
\begin{figure}
     \vspace{3.85in}     
     \includegraphics{#1} 
	\renewcommand{\lastlabel}{#2}
	\caption
}%
{\label{\lastlabel}\end{figure}} 

{
\begin{figure}[h]
     \vspace{3.5in} 
     \includegraphics{#1} 
	\renewcommand{\lastlabel}{#2}
	\caption
}%
{\label{\lastlabel}\end{figure}} 
{
\begin{figure}[t]
     \vspace{3.5in} 
     \includegraphics{#1} 
	\renewcommand{\lastlabel}{#2}
	\caption
}%
{\label{\lastlabel}\end{figure}} 
{
\begin{figure}[b]
     \vspace{3.5in} 
     \includegraphics{#1} 
	\renewcommand{\lastlabel}{#2}
	\caption
}%
{\label{\lastlabel}\end{figure}} 
{
\begin{figure}[p]
     \vspace{3.5in} 
     \includegraphics{#1} 
	\renewcommand{\lastlabel}{#2}
	\caption
}%
{\label{\lastlabel}\end{figure}} 

{
\begin{figure}
     \vspace{5.05in} 
     \includegraphics{#1} 
	\renewcommand{\lastlabel}{#2}
	\caption
}%
{\label{\lastlabel}\end{figure}} 

{
\begin{figure}
     \vspace{3.85in}     
     \includegraphics{#1} 
	\renewcommand{\lastlabel}{#2}
	\caption
}%
{\label{\lastlabel}\end{figure}} 

{
\begin{figure}[h]
     \vspace{3.85in}     
     \includegraphics{#1} 
	\renewcommand{\lastlabel}{#2}
	\caption
}%
{\label{\lastlabel}\end{figure}} 

{
\begin{figure}
     \vspace{2.5in} 
     \includegraphics{#1} 
	\renewcommand{\lastlabel}{#2}
	\caption
}%
{\label{\lastlabel}\end{figure}} 

{
\begin{figure}
     \vspace{2.5in} 
     \includegraphics{#1} 
	\renewcommand{\lastlabel}{#2}
	\caption
}%
{\label{\lastlabel}\end{figure}} 

{
\begin{figure}
     \vspace{6.5in} 
     \includegraphics{#1} 
	\renewcommand{\lastlabel}{#2}
	\caption
}%
{\label{\lastlabel}\end{figure}} 

{
\begin{figure}
     \vspace{3.0in} 
     \includegraphics{#1} 
	Figure #2
}%
{\end{figure}} 

{
\begin{figure}
     \vspace{5.25in} 
     \includegraphics{#1} 
	Figure #2
}%
{\end{figure}}